# A Study of Apparent Symmetry Breakdown in Perovskite Oxide-based Symmetric RRAM Devices

X. Chen, J. Strozier, N.J. Wu[a], A. Ignatiev, Y.B. Nian

Texas Center for Superconductivity and Advanced Materials

University of Houston, Houston, TX 77204

Monday, November 29, 2004

## Abstract

A new model of a symmetric two-terminal non-volatile RRAM device based on Perovskite oxide thin film materials, specifically $Pr_{1-x}Ca_xMnO_3$ (PCMO), is proposed and analyzed. The model consists of two identical half-parts, which are completely characterized by the same resistance verses pulse voltage hysteresis loop, connected together in series.  Even though the modeled device is physically symmetric with respect to the direction of current, it is found to exhibit switching of the resistance with the application of voltage pulses of sufficient amplitude and of different polarities. The apparent breakdown of parity conservation of the device is attributed to changes in resistance of the active material layer near the electrodes during switching.  Thus the switching is history dependent, a feature that can be very useful for the construction of real non-volatile memory devices. An actual symmetric device, not previously reported in the literature and based on the proposed model, is fabricated in the PCMO material system.  Measurements of the resistance of this new device generated an experimental hysteresis curve that matches well the calculated hysteresis curve of the model, thus confirming the features predicated by the new symmetric model.

[a] Electronic mail: naijwu@uh.edu

The recent observation of the electrical pulse induced resistance change effect (EPIR) [1] in Perovskite oxide thin films at room temperature has drawn much attention [2-5]. An oxide film such as $Pr_{1-x}Ca_xMnO_3$ (PCMO) sandwiched between two electrodes forms a two terminal EPIR device. The resistance of the device can be switched down by a short (~10 ns) electric pulse of one polarity (one direction of current) and switched up by a short pulse of the opposite polarity (other direction of current). Such simple structured resistive switching devices have important application in novel non-volatile high density memory devices. [2, 4, 6] The basic mechanism responsible for the EPIR effect is still evading researchers, with projections of both bulk [1, 7-10] and interface [11] properties responsible for the origin of the effect. Initial studies on fabricated devices [1] recognized the possibility of non-symmetry in the device structure since in general the top and bottom electrodes were of different sizes and different materials. This asymmetry due to electrode/electrode-region variability has been recently incorporated as a requirement for reversible switching in the experimental work of Aoyama, et al, [10] and in the theoretical studies of Rosenberg et al, [8] who generated different top and bottom metallic non-percolating domains (related to the electrodes) as the basis for the effect.

However, it has also been observed that an apparently symmetric EPIR device (a PCMO thin film with two equal sized gold electrodes) can also show switching behavior [12]. In this system with device asymmetry removed, it appears that conservation of parity may be violated, and hence models other than one incorporating an intrinsic bulk material effect have been proposed. [11] In the model proposed here, we focus on a symmetrically structured device, and resolve this resistance switching paradox for a symmetric system.

We describe in this letter a device resistance switching model carefully constructed to be symmetric with two identical half-parts. The two parts are not only structurally identical, but have the same resistance switching characteristics as given by a hysteresis curve of resistance verses voltage. Moreover, they are also in the same initial resistive states. The device is subjected to electrical pulses of different pulse amplitudes and polarity, the resistance switching behavior is studied, and the apparent parity breakdown is discussed. Finally, an actual symmetrically structured EPIR device based on the model is constructed and experimentally tested for the features predicted by the model.

Fig. 1 shows the schematic of a symmetric device consisting of two identical half-parts as shown in Fig. 1(a). The left part of the device consists a top terminal electrode A in contact with a functional oxide switching layer, which has been fabricated on a bottom electrode. A reference electrode O is connected to the bottom electrode. The structurally identical right half-part of the device has a top terminal electrode B.

Either of the two half-devices can be viewed as a two-electrode device, which is not symmetric. Thus, there is no parity paradox to resolve in understanding the resistance switching characteristics of the half device. We assume the resistance switching hysteresis behavior of the left half-device shown in Fig. 1(b). Note that the resistance hysteresis loop of the half-device is "non-symmetrical" in the sense that the transition voltages are different for the two pulse polarities. This non-symmetric half-device loop assumption is based on our experimental observations in actual devices, where the transition voltages for switching resistance up and down in a hysteresis loop are usually not the same. We wish to point out that as long as the two halves are identical images to each other, half-device loop asymmetry does not affect total device symmetry. As we will show later, the half-device hysteresis loop non-symmetry is not a required condition for the total device to be switchable.

For switching the left half-device, we chose the direction from electrode A to O as the positive pulse direction, and in modeling the application of short electrical pulses of a given voltage across the two electrodes of the half-device, we obtain the resistance hysteresis curve shown in Fig 1(b). When the half-device is in the high resistance state of a modeled 10Ω, a small pulse voltage does not change the resistance. However when the pulse voltage is increased to over 5V, the resistance begins to drop. At 6V, the resistance drops to the modeled low state of 5Ω, and saturates at that value when pulses of higher voltage are applied. When the half-device is in the low state, a positive pulse or a small negative pulse does not change the resistance. When the negative pulse height increases to a value of –7V, the resistance starts to increase, and reaches the high resistance state of 10 Ω after a pulse of –8V is applied, saturating at the 10Ω value with increasing negative pulse voltage. The cycle is then repeated as the pulse voltage first decreases in negative amplitude and then moves towards more positive values. Note that in all cases referred to in this paper, the resistance is measured or calculated after the pulse, using voltages much smaller than those used for switching.

The right half-device is identical to the left half-device, and by setting the positive direction in the B-O direction for the right half-device, the same hysteresis loop as shown in Fig 1(b) will decide the right half-device behavior.

Fig. 2(a) shows a computer calculation of the hysteresis curve of resistance vs. pulse voltage for the whole two terminal A-B device by using the hysteresis behavior modeled above for each of the half-devices. In calculating the applied pulse voltage that causes the switching, we calculate the voltage across each of the half-devices taking into account the current resistance of each of the two parts. We also assume that in the switching region, the half-device resistance varies linearly with the pulse voltage drop across it, and is a first order transition. That is, switching does not take place in the switching region for decreasing (absolute values) of the pulse voltage. Obviously history is important in determining the voltage across each part at each point in the hysteresis curve, and this fact is built into the computer calculation.

We begin the calculation with both half-devices in the high resistance state. Thus the resistance of the whole device is 20 Ω (point 1 in Fig. 2(a)). Pulses are applied across the whole device (A-B) with increasing value. At +10V (point 2), the strict symmetry of the whole device breaks down as the resistance of the left part, $R_{A-O}$ begins to decrease, while the resistance of the right part, $R_{B-O}$ remains the same. Further increase of the pulse voltage results in resistance saturation (starting at +18V, point 3) at 15 Ω (the left part is at 5 Ω and the right part at 10 Ω). If the pulse voltage is now decreased; nothing happens until the pulse voltage reaches a value of -7.5 V (point 4). At this point the resistance of the right part begins to decrease reaching a value of 5 Ω at a pulse voltage of -12 V (point 5). Now both parts are in their low resistance state, so that the resistance of the whole device is 10 Ω. Further decrease in pulse voltage results in no change until the pulse voltage reaches -14 V (point 6) at which point the resistance rises rapidly to 15 Ω as the left part switches back up to 10 Ω (point 7) and remains there as the pulse voltage is further decreased. Reversal of the pulse potential follows a mirror resistance curve with a decrease beginning at +7.5 V (point 8), reaching a low value of 10 Ω at +12V (point 9), and then starting to rapidly rise at +14V to a value of 15 Ω (point 11). Continued increase in the pulse voltage results in no change in the resistance. The hysteresis curve saturates at 15 Ω for both large positive pulse voltages (greater than +14V) and large negative pulse voltages (less than -14V). In the cycle, points 3-11 are repeatable.

The reason for the apparent violation of parity in the whole device is that in the middle resistance state (15 Ω), the two half-parts are in different resistance states. Thus the whole device is no longer symmetrical. We also note that after the initial drop in resistance (points 2-3) the resistance of the whole device saturates at 15 Ω for voltages less than -14V or voltages greater than +14V. Also, depending on history, the whole device switches down in resistance for voltages between -7.5 to -14V or +7.5 to +14V. The resistance change of each of the two-half devices as function of pulse voltage $V_{A-B}$ is shown in Fig 2(b), which can be constructed directly from Fig 2(a).

We furthermore calculate directly the switching characteristics of the model, which is shown in Fig 2(c). We first use a large positive voltage pulse across A-B (first training) to set the left side to the high resistance state and the right side to the low resistance state (symmetry breakdown state), and then we subject the device to alternative ±16V pulses. We note that the resistance R as measured across A-B switches back and forth between 10Ω and 15Ω as pulse voltage, V goes from +16 volts to -13 volts. We can then apply a large negative voltage pulse across A-B (second training) and then again apply alternative -16V and +13V pulses to the device with the device resistance switching back and forth between 10Ω and 15Ω. However, the device now switches to the low resistance state with -16 V pulses and to the high resistance state with +13V pulses. This occurs because the apparent non-symmetry created in the back and forth switching is not intrinsic to the device, but only results from the asymmetric resistance of the two half-parts as a result of prior training.

From Fig 2(a), one can construct a partial hysteresis curve for the whole device in the range defined by pulse voltages greater than -14 volts. This partial hysteresis curve is given in Fig 2(d). Note that the resistance saturates at 15Ω (the high state) for high positive voltages, but does not saturate at 10Ω (the low state) for arbitrarily high negative voltages. In particular, for pulse voltages less than -14 volts the resistance switches up to 15 Ω and saturates at that value as the pulse voltage is further decreased.

To show under certain conditions that a device with a "symmetrical" half-device hysteresis loop could switch, we examine a device where the half-device hysteresis loop has a "symmetric" curve, i.e., we changed the starting negative switch voltage to –5V and the saturating negative switching voltage to –6V, so that the starting and saturating negative switch voltages have the same absolute values as the corresponding positive switch voltages. The high state is 10Ω, the low state is 5Ω. The hysteresis loop of this device in the full voltage range is shown in Fig. 3 and is very similar to that of Fig 2(a).

The symmetrical device discussed here is not the same as the EPIR switching devices now commonly studied [1, 10-11]. These EPIR devices are asymmetrical devices, where asymmetry factors exist such as different material and size for top and bottom electrodes [1], point-to-point variation on a sample [11], or two top electrodes made of two different materials [10]. Because those devices have built-in asymmetry, there is no real apparent parity violation problem. One should not expect those devices to show symmetrical response to pulses. The model device we presented here is a carefully constructed symmetrical device, in which the half-devices are not necessary to be symmetric in their hysteresis loops, but the whole device is perfectly symmetric. It has some characteristics similar to the commonly studied EPIR devices, such as a hysteresis loop that indicates resistance switching back and forth under pulses of different polarities [12]. However, the model also shows some new characteristics not yet realized in existing EPIR devices, such as a symmetric stabilized hysteresis curve under the full pulse voltage range. The study not only provides the possibility to develop a new two terminal symmetrical non-volatile memory device, but also a new three terminal memory multi state memory devices if we use the reference electrode as a device terminal.

To verify these model results, an actual symmetrically structured EPIR device was constructed and tested. Each half-part of the actual device consists of a layered structure of Ag/PCMO/YBa$_2$Cu$_3$O$_{7-x}$ on a LaAlO$_3$ substrate structured as described by Liu et al [1]. A silver contact was made to the exposed YBa$_2$Cu$_3$O$_{7-x}$ bottom electrode surface. The two half-devices were first tested to confirm that they are very close in switching behavior to each other. Then the whole device was assembled and put under test. The results are shown in Fig 4(a) where the switching hysteresis loop of the full device is similar to the curves in Fig 2(a) and Fig 3 with the "table with legs" features. The hysteresis loops of the half-devices during this test (Fig 4(b)) are also in good agreement with model analysis shown in Fig 2(b). Thus, the data from an actual device demonstrate good symmetry within experimental error and support the newly defined model of resistance switching in a thin film oxide device.

The study presented here not only gives some understanding of the apparent parity violation concern in a structurally resistive switching symmetrical device, but also suggests the fabrication and application of a new type resistance switching device. Instead of requiring the two terminals to be structurally different, [10] one can keep them structurally symmetrical while using pulse training to introduce resistance asymmetry and thereby create a switching device. The model results are confirmed by the experimental results on a fabricated symmetric device.

## ACKNOWLEDGMENTS

The assistance from C. Papagianni, Z.W. Xing, L. Smith, A. Zomorodian, Y. Song is very much appreciated. This research was partially supported by NASA, the State of Texas through TcSAM, Sharp Laboratories of America, and the R. A. Welch Foundation.

**FIGURE CAPTION**

Fig 1. The modeled symmetric device a), and b) the switching hysteresis loop of the half device A-O (V=$V_{A-O}$, R=$R_{A-O}$), which is identical to the loop of half device B-O (V=$V_{B-O}$, R=$R_{B-O}$).

Fig 2. Computer calculations on the structurally symmetric whole device A-B, demonstrating apparent symmetry breakdown. a) The symmetric (lower part) hysteresis curve of resistance verses pulse voltage for the whole device A-B; b) The hysteresis curves of the half devices; c) Low-high state switching behavior of A-B based on training with a single voltage pulse: initial training +20V, second training -20V; d) the asymmetric hysteresis curve of A-B (the apparent asymmetry can be reversed by training).

Fig 3. The switching of a full device with a "symmetrical" half device loop

Fig 4. The experimental hysteresis loop of an actual EPIR device with a symmetric structure. a) The resistance verses pulse voltage for the total device; b) The resistance verses pulse voltage of each of the two half devices.

**FIGURES**

Xin Chen, et al. (A Study of Apparent…)

**Fig 1.**

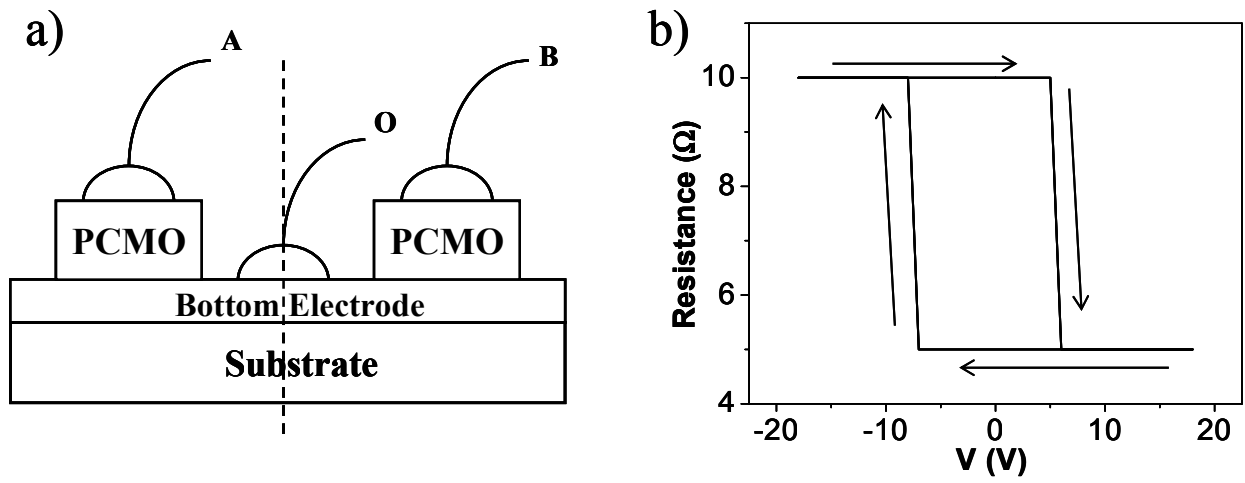

**Fig 2.**

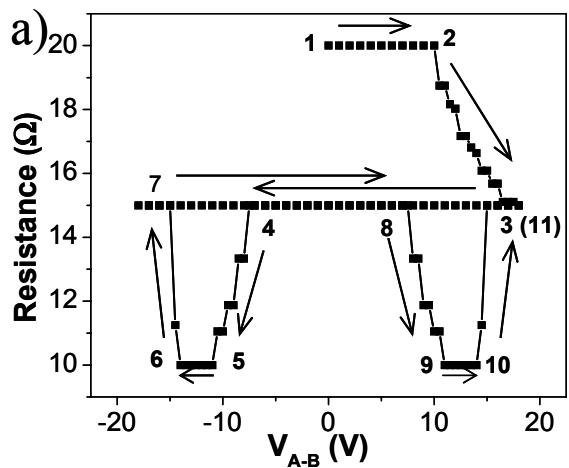
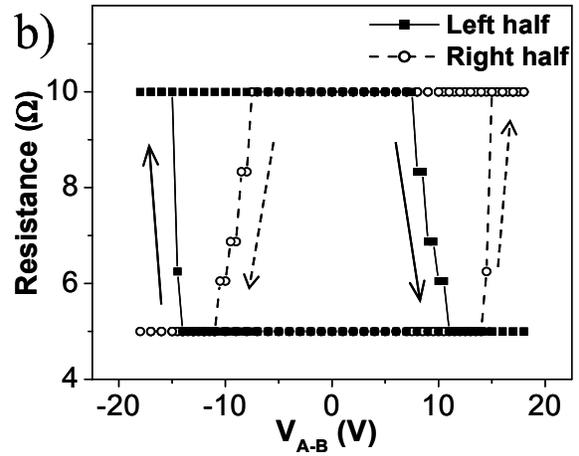
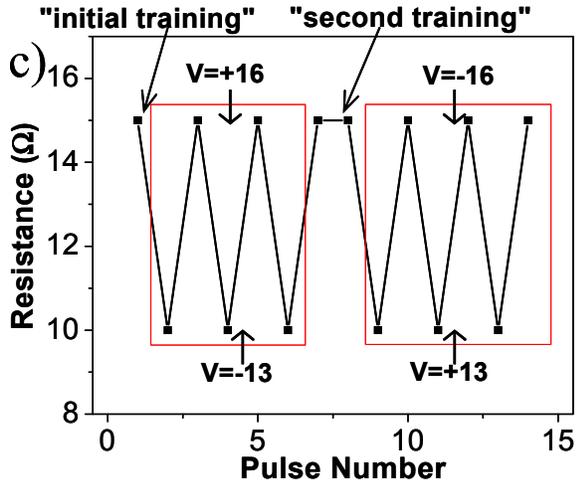
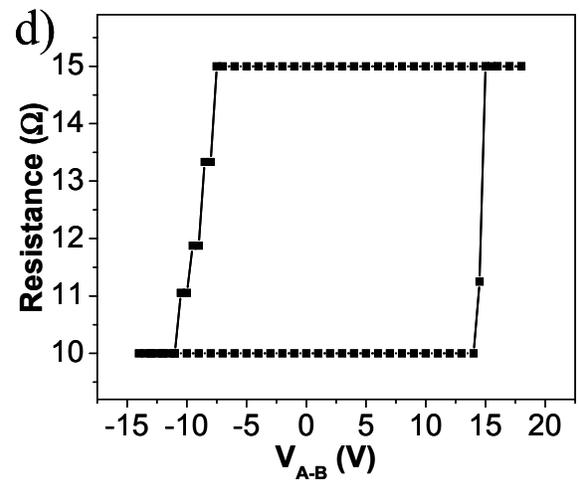

**Fig 3.**

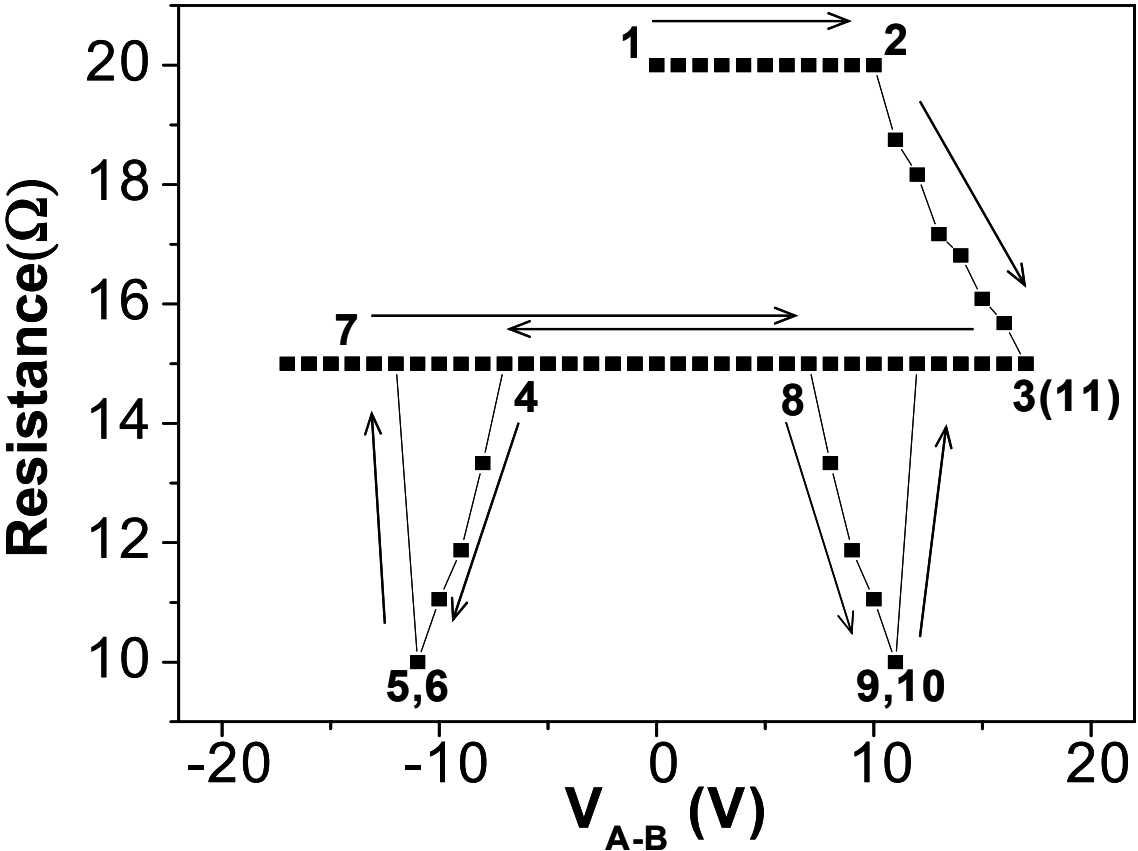

**Fig 4.**

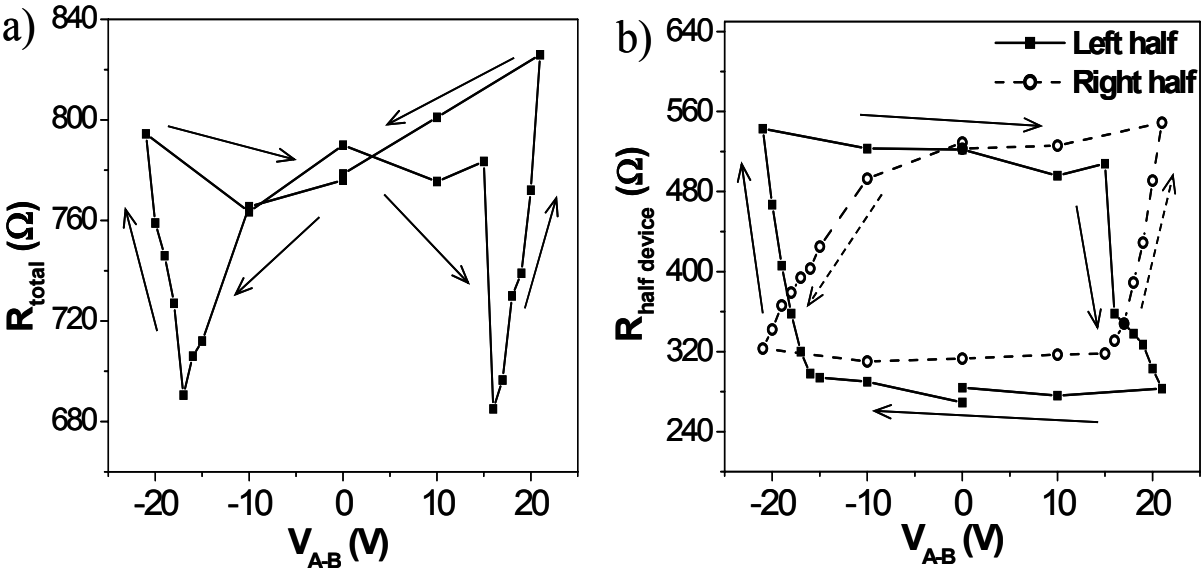